\documentclass[]{article}
\usepackage{authblk}

\usepackage{booktabs} 
\usepackage[utf8]{inputenc} 
\usepackage[T1]{fontenc}    
\usepackage{hyperref}       
\usepackage{url}            
\usepackage{amsfonts}       
\usepackage{nicefrac}       
\usepackage{microtype}      
\usepackage{fullpage}
\usepackage{longtable}
\usepackage{adjustbox}
\usepackage{amsmath}
\usepackage{graphicx}
\usepackage{paralist}
\usepackage{mathtools}
\usepackage{xfrac}
\usepackage{bm}
\usepackage{bbm}
\usepackage{caption}
\usepackage{subcaption}
\usepackage{multirow}
\usepackage[inline]{enumitem}
\usepackage[title]{appendix}
\newcommand{\explain}[2]{\underset{\mathclap{\overset{\uparrow}{#2}}}{#1}}
\newcommand{\explainup}[2]{\overset{\mathclap{\underset{\downarrow}{#2}}}{#1}}
\newcommand{\norm}[1]{\left\lVert#1\right\rVert}

\newcommand{\xhdr}[1]{\vspace{1.2mm}\noindent{{\bf #1.}}}

\begin{document}
\title{Homogeneity-Based Transmissive Process To Model True and False News in Social Networks}

\author[1]{Jooyeon Kim}
\author[1]{Dongkwan Kim}
\author[1]{Alice Oh}

\affil[1]{KAIST, jooyeon.kim@kaist.ac.kr, dongkwan.kim@kaist.ac.kr, alice.oh@kaist.edu}

\date{}

\maketitle
\begin{abstract}
An overwhelming number of true and false news stories are posted and shared in social networks, and users diffuse the stories based on multiple factors. Diffusion of news stories from one user to another depends not only on the stories' content and the genuineness but also on the alignment of the topical interests between the users. In this paper, we propose a novel Bayesian nonparametric model that incorporates \textit{homogeneity} of news stories as the key component that regulates the topical similarity between the posting and sharing users' topical interests. Our model extends hierarchical Dirichlet process to model the topics of the news stories and incorporates Bayesian Gaussian process latent variable model to discover the homogeneity values. We train our model on a real-world social network dataset and find homogeneity values of news stories that strongly relate to their labels of genuineness and their contents. Finally, we show that the supervised version of our model predicts the labels of news stories better than the state-of-the-art neural network and Bayesian models.
\end{abstract} \vspace{-2mm}
%
%
\maketitle
\section{Introduction}
In recent years, users in social media and microblogging platforms have played major roles in disseminating online content and news stories~\cite{vosoughi2018spread,holcomb2013news}.
While some are based on factual information, a non-negligible portion of the news contains false information that often stir unnecessary disputes, sway people's opinions, give rise to political polarization, and even cause financial losses in the stock market ~\cite{kumar2018false,bessi2015science,mocanu2015collective,bessi2014economy,rapoza_2017}.

In reaction to this, meaningful academic progress has been made to uncover the spreading pattern of true and false news in social networks~\cite{tambuscio2015fact,wu2018tracing,kwon2013prominent}, detect misinformation from content, network, and temporal information~\cite{kwon2017rumor,ma2015detect,ma2017detect,ma2018rumor,zhao2015enquiring,rashkin2017truth}, early-fact-check and curtail the spread of falsity~\cite{kim2018leveraging,friggeri2014rumor,tschiatschek2018fake,balmau2018limiting}, build fact-checking systems~\cite{nguyen2018believe,vo2018rise}, and investigate the potential intervention of user biases in the process of fact-checking~\cite{babaei2018analysing}. 
Also, recent work by Vosoughi \textit{et al}. conducts an in-depth investigation into the differences between true and false information in the aspects of \textit{topics}, \textit{cascade patterns} and sizes, diffusion speeds, emotion, and sentiment~\cite{vosoughi2018spread}. Another recent paper by Del Vicario \textit{et al}. suggests that \textit{homogeneity} among users is the main driving factor of the dissemination of both true and false news~\cite{del2016spreading}.

Motivated by such progress, we propose Homogeneity-Based Transmissive Process (HBTP), a Bayesian nonparametric model that captures the complex interplay between the textual content, topics of true and false news, and the topical interests of users who share the news in social networks. 
Specifically, our model is operationalized as follows:
\begin{enumerate}
    \item Each user is assigned \textit{user interest} which is computed from the collection of content information of news stories that the user shares.
    \item Each news story is assigned a \textit{homogeneity index} which regulates the degree of homogeneity. We define this document-centric (rather than user-centric) homogeneity index to be the degree of uniformness of user interests among users who share the story. \vspace{-2mm}
\end{enumerate}
Under our modeling assumption, a user's topical interest \textit{transmits} to another user through the process of them co-sharing a news story in a social network, and the degree of transmission between users depends on the homogeneity index of the news story that the users co-share.
Thus, suppose there are two news stories, one with a high homogeneity index and the other with a low index.
The news story with the high index will propagate through a highly biased subgroup within a social network, 
while propagation of the news story with low homogeneity index will look more like a random walk.

To formulate such modeling assumptions we design our HBTP model as follows:
\begin{enumerate}
    \item[(a)] Within the nonparametric topic modeling framework~\cite{teh2005sharing}, we model the content of news stories from the topical interests of users who participate in diffusing the news story.
    \item[(b)] We formulate each user's topical interest as a single probability measure in two-layer Dirichlet processes~\cite{ferguson1973bayesian} and allow it to be drawn not only from the single, upper-level probability measure but also from multiple probability measures of the preceding users.
    Also, we adopt the Gamma process construction of HDP~\cite{paisley2012discrete} with a modification to incorporate the homogeneity index and regulate the homogeneity among user interests.
    \item[(c)] We combine nonparametric topic models with Bayesian Gaussian process latent variable models (Bayesian GP-LVMs)~\cite{titsias2010bayesian} to infer the homogeneity index of a news story from the users' and the stories' topics in a highly nonlinear fashion.
\end{enumerate}
Since both the content modeling and homogeneity discovery part of our modeling fall into the Bayesian nonparametric framework, the posterior inference of our model can be done using the variational Bayes methods~\cite{jordan1999introduction}.
We train our model on a real-world Twitter dataset~\cite{liu2015real,ma2016detecting,ma2017detect} that consists of user posting and sharing with true and false news stories.
Our main contributions and the findings through the experiments are as follows: 
\begin{enumerate}
    \item[I.] We develop HBTP that jointly models contents of news stories and user sharing events.
    Our model discovers the latent topics, topical interests of users, and homogeneity indices of news stories; these mutually reshape one another in the joint modeling.
    \item[II.] We find that the homogeneity indices that our model discovered significantly varies according to the labels of genuineness of the stories.
    \item[III.] We develop a supervised extension of HBTP (sHBTP) that incorporates the genuineness of news stories in the training of the model, and conduct a classification task of predicting the genuineness of a news story in the test set.
    We compare our prediction results with state-of-the-art fake news detection models and with other recently proposed supervised topic models and show that our model outperforms the comparison models in most of the evaluation metrics.
    \item[IV.] We make our code and data (Twitter data augmented with the news articles that the tweets refer to) publicly available for future research (\url{https://github.com/todoaskit/HBTP}).
\end{enumerate}
\xhdr{Related work}
By incorporating the label information of the news stories, the supervised formulation of our model performs the classification task of predicting the labels of the news stories. There are feature-based methods that leverage text, social network, temporal traces and propagation models to classify true and false news in supervised fashion~\cite{ma2017detect,qazvinian2011rumor,gupta2013faking,kumar2016disinformation,yang2012automatic,wu2015false,kwon2017rumor,kwon2013prominent}. Also, another line of research focuses on devising algorithms that mitigate false news and their diffusion in social networks~\cite{budak2011limiting,tripathy2010study,nguyen2012containment,kim2018leveraging,farajtabar2017fake}.
%
%
%
%
%
%
%
%
%

From the modeling aspect, there are upstream models in parametric topic modeling literature that generate the topic indices in a document from the topic distribution of a set of entities~\cite{rosen2004author,mimno2012topic,dietz2007unsupervised,mccallum2005topic,mimno2007expertise}. HBTP falls into this category in a way that a set of users who share a news story participate in modeling its content. 
%
%
%
%
Also, to generate both homogeneity values and labels of news stories, we incorporate Bayesian GP-LVM~\cite{titsias2010bayesian,damianou2016variational} in our model in the nonparametric topic modeling framework. Bayesian GP-LVM extends GP-LVM~\cite{lawrence2004gaussian,lawrence2005probabilistic} by incorporating additional priors and conducting posterior inference using variational methods~\cite{jordan1999introduction,wainwright2008graphical}. Kandemir \textit{et al}. recently proposed GPSTM as a joint model between LDA and Bayesian GP-LVM and demonstrate the performance gain of jointly modeling LDA and GPLVM over linear classifiers~\cite{kandemir2017supervising}.
%
%
%
%
Lastly, while our model focus on the joint modeling of text and users' sharing patterns~\cite{kim2017joint}, textual data has also been aligned with users' temporal traces to model diffusion pattern and text~\cite{du2013uncover,he2015hawkestopic}, to cluster the documents in both discrete~\cite{ahmed2010timeline} and continuous~\cite{du2015dirichlet,wang2014mmrate,mavroforakis2016modeling} time domains. These models can be complementary to our model and be co-trained to increase the predictive power in detecting false news. \vspace{-2mm}
%
%
%
\section{Background on the Gamma process construction of HDP} 
Hierarchical Dirichlet process (HDP) for modeling document collections is a two-level Dirichlet process (DP)~\cite{ferguson1973bayesian} that functions as a nonparametric Bayesian prior for mixed membership models~\cite{teh2005sharing}. Being nonparametric, we do not have to specify the number of topics a priori, but instead, it can be inferred from the data~\cite{teh2005sharing}. 

In the HDP topic model, the random probability measure drawn from the first-level Dirichlet process becomes the base measure for the second-level Dirichlet processes: \vspace{-2mm}
\begin{equation*}
    G_0 \sim \text{DP}(\alpha, H), \quad G_d \overset{iid}{\sim} \text{DP}(\beta, G_0), 
\end{equation*} 
where $d$ is the document index, $H$ is the base measure, and $\alpha, \beta$ are the first and the second-level DP concentration parameters respectively. 
%
%
$H$ is a Dirichlet distribution on the vocabulary simplex, and the atoms of $G_0$ are an infinite set of word-topic distributions drawn from $H$. 

The first- and the second-level Dirichlet processes yield random discrete probability measures $G_0 = \sum_{k = 1}^{\infty}{p_k \delta_{\phi_k}}$ and $G_d = \sum_{k = 1}^{\infty}{p^{(d)}_k \delta_{\phi_k}}$. Here, the weights $\{ p_k \}_{k = 1}^{\infty}$ and $\{ p^{(d)}_k \}_{k = 1}^{\infty}$ depend on the corresponding concentration parameters $\alpha$ and $\beta$ and atoms \smash{$\{ \phi_k \}_{k = 1}^{\infty}$} are drawn from the base probability measure: \smash{$\phi_k \overset{iid}{\sim} H$}. Finally, each second-level $G_d$ generates its \smash{$n^{th}$} associated topic-indicator variable \smash{$z^{(d)}_n$} and each word \smash{$x^{(d)}_n$} is drawn from \smash{$\phi_{z^{(d)}_n}$} for $N_d$ times as follows: \vspace{-2mm}
\begin{equation*}
    z^{(d)}_n | G_d \overset{iid}{\sim} G_d, \quad  x^{(d)}_n | z^{(d)}_n \overset{iid}{\sim} \text{Multinomial}(\phi_{z^{(d)}_n}). 
\end{equation*}

Following the work of Paisley \textit{et al.}~\cite{paisley2012discrete}, we constructively define the first and the second-level Dirichlet processes using the stick-breaking process~\cite{sethuraman1994constructive} and normalized Gamma processes~\cite{paisley2012discrete}. For the first-level $G_0$, we represent the weights $\{ p_k \}_{k = 1}^{\infty}$ as \vspace{-2mm}
\begin{equation*}
    p_k = V_k \prod_{j = 1}^{j < k}{(1- V_k)}, \quad V_k \overset{iid}{\sim} \text{Beta}(1, \alpha). 
\end{equation*}
The second-level probability measures $G_d$ are conditionally distributed on the first-level random probability measure $G_0$, and the corresponding weights \smash{$\{ p^{(d)}_k \}_{k = 1}^{\infty}$} can be represented as \vspace{-2mm}
\begin{equation*}
    p^{(d)}_k = \frac{\pi_k^{(d)}}{ \sum^{\infty}_{j = 1} \pi_j^{(d)} }, \quad \pi_k^{(d)} | G_0 \overset{ind}{\sim} \text{Gamma}(\beta p_k, 1).
\end{equation*}
\section{Homogeneity-Based Transmissive Process}
In this section, we introduce the modeling scheme of the homogeneity based transmissive process. 
We start by specifying different kinds of data structure the model works on. 
Next, we describe the generative process of the first- and second-level Dirichlet processes,
explain the transmissive property when generating the second-level probability measures and explain how the homogeneity indices of news stories function in regulating those probability measures.
We then explain how we generate the content of a news story as a mixture of users' probability measures
and illustrate how we draw homogeneity indices of news stories.
Lastly, we discuss the parametric and supervised formulations of our model.
\vspace{-2mm}
\subsection{Event representation}
In social networks, users introduce news stories and these stories are propagated throughout the network by means of sharing (\textit{e.g.}, tweets and retweets). We first represent such tweet and retweet events in a social network as a triplet \vspace{-2mm}
\begin{equation}
    e := (\explain{u}{\text{\vphantom{k}user}}, \quad \explainup{a}{\text{preceding user}}, \quad \explain{s}{\text{story}}).\nonumber 
\end{equation}
Here, we denote a set of events as $\mathbb{E} \ni e$.
The first item $u$ is a user who created the tweet/retweet, and the second item $a$ is a user who created the preceding tweet if the event is a retweet. 
%
%
%
We assume that there are $M$ users and denote a set of users as $\mathbb{U} \ni u, a$. 
The last item $s$ is a news story. We represent each news story as a triplet\vspace{-2mm}
\begin{equation}
    s := (\explain{d}{\text{\vphantom{k}content}}, \quad \explainup{h}{\text{homogeneity index}}, \quad \explain{l}{\text{label}}), \nonumber
\end{equation}
where $d$ is a bag-of-words representation of content of the story, $h$ is a homogeneity index, and $l$ is the label of the story, \textit{e.g.}, true or false. We remark that the homogeneity index $h$ is hidden and will be uncovered using our model. There are $L$ stories and the set of stories is denoted as $\mathbb{S} \ni s$.\vspace{-2mm}

\subsection{Modeling the first and the second-level DPs}

Our model inherits the two-level DP construction of the HDP, and the first-level DP formulates identically to that of the HDP. We draw a probability measure $G_0$ from DP with the base measure $H$ and the concentration parameter $\alpha$. Using the stick-breaking construction~\cite{sethuraman1994constructive}, we represent the first-level probability measure as $G_0 = \sum^{\infty}_{k=1}V_k \prod^{j=1}_{j<k}(1 - V_j)\delta_{\phi_k}$, $V_k \overset{iid}{\sim}\text{Beta}(1, \alpha)$ and $\phi_k \overset{iid}{\sim}H$.

\xhdr{Transmissive modeling of user DPs} 
For the second-level, we endow each user in a social network, instead of a news story, with a single DP.
The key idea is that each user's probability measure is transmitted to the users who retweeted her (re)tweet, and the degree of transmission, \textit{i.e.}, the similarity between the two users' probability measure, depends on the homogeneity index of the news story the users are sharing.
To formalize this, for each user $u$, we first construct $\mathbb{E}_u$ which is a set of events $e$ whose first element is $u$. 
%
%
From $\mathbb{E}_u$ we retrieve $\mathbb{A}_u \ni v$, a set of users preceding $u$ and denote the set of news stories that are shared by $u$ and $v$ as $s^{(u, v)}$.

%
When a user $u$ has no predecessor, \textit{i.e.,} $\mathbb{A}_u = \emptyset$, the user's probability measure $G_u$ is drawn from\vspace{-2mm}
\begin{equation*}
    G_u = \sum^{\infty}_{k = 1} \frac{\pi_k^{(u)}}{ \sum^{\infty}_{j = 1} \pi_j^{(u)} } \delta_{\phi_k} \sim \text{DP}(\beta, G_0), \quad \pi_k^{(u)} \sim \text{Gamma}(\beta p_k, 1). \vspace{-2mm}
\end{equation*}
Recall that $p_k = V_k \prod^{k-1}_{j=1}(1 - V_j)$ is the weight of the $k^{th}$ atom of the first-level probability measure.

If the user $u$ has preceding users, for each $v \in \mathbb{A}$ we draw $G_u^{(v)}$, which is user $u$'s probability measure \textit{transmitted} by the preceding user $v$. Denoting the homogeneity index of story $s^{(u,v)}$ as $h^{(u,v)}$, $G_u^{(v)}$ is drawn from the DP with the base measure $G_v$ and the concentration parameter $\beta \exp{(h^{(u, v)})}$ as follows: \vspace{-1mm}
\begin{align}
    G_u^{(v)} &= \sum^{\infty}_{k = 1} \frac{\pi_k^{(u, v)}}{ \sum^{\infty}_{j = 1} \pi_j^{(u, v)} } \delta_{\phi_k} \sim \text{DP}(\beta  \mathrm{e}^{h^{(u, v)}}, G_v), \nonumber \\
    \pi_k^{(u, v)} &\sim \text{Gamma}(\beta \mathrm{e}^{h^{(u, v)}} p^{(v)}_k, 1), \vspace{-2mm} 
    \label{g_u^a}
\end{align}
where $p_k^{(v)} = \sum^{\infty}_{k = 1} \tfrac{\pi_k^{(v)}}{ \sum^{\infty}_{j = 1} \pi_j^{(v)} }$ is the weight of the $k^{th}$ atom of the preceding user $v$'s probability measure. 
Here, we posit that there is no circular retweet among users.
Finally, denoting the weight of the $k^{th}$ atom of $G_u^{(v)}$ as $p_k^{(u, v)}$, we constructively define user $u$'s probability measure as \vspace{-2mm}
\begin{equation*}
    G_u := \sum^{\infty}_{k=1} \frac{\sum_{v \in \mathbb{A}_u} p_k^{(u, v)}}{ \left\vert \mathbb{A}_u \right \vert } \delta_{\phi_k}.
\end{equation*}
Note that $G_u$ derived from multiple $G_u^{(v)}$s is still a probability measure as it meets the countable additivity property and the weights of the atoms lie in the unit interval~\cite{roussas2014introduction}. 

\xhdr{Remark 1} 
We modify the shape parameter of $\text{Gamma}(x \vert a, b) = \tfrac{b^a}{\Gamma{(a)}} x^{a - 1} \exp{(- b x)}$ in equation~\ref{g_u^a} to incorporate the effect of homogeneity of news stories on the transmission of the probability measure.
This approach differs from past approaches which modify the rate parameter $b$ to achieve topic-wise re-scaling~\cite{paisley2012discrete,kim2017hierarchical}.
In our model, changing the rate parameter would be ineffective because the expected impact of the homogeneity index is identical for all topics, and scaling the rate parameter using the same multiplier for all topics yields identical outcome without the multiplier in the Gamma process after the normalization step. 
For this particular reason, the rate parameters in the Gamma processes disappear when used to represent DP~\cite{devroye1986sample}.
Conversely, by scaling the shape parameter by, say, $\tau$, the mean of the normalized Gamma process remains unchanged while the variance is reduced by $1 / \tau$. Therefore, when a probability measure of a user is \textit{transmitted} by a news story with a high homogeneity index, the probability measure of the user will be more similar to a preceding user's probability measure.
Finally, note that since the rate parameter remains unchanged, our formulation of the second-level probability measures can be represented using the stick-breaking process.
We use the normalized Gamma process representation to enjoy the merit of generating $\pi_k$ independently, which will be useful for approximate posterior inference. \vspace{-2mm}
\subsection{Modeling news stories using mixtures of user DPs}
We model the contents of a news story $s$ using probability measures of users who participated in propagating the story. 
There are multiple previous parametric topic models that adopt this \textit{upstream} approach to model a document using probability measures from multiple sources~\cite{rosen2004author,mimno2012topic,dietz2007unsupervised,mccallum2005topic,mimno2007expertise}. 
Here, using $\mathbb{E}$, we define $\mathbb{U}_s$ as the set of users who propagated the story $s$.
Then, following the work of Rosen-Zvi \textit{et al.}~\cite{rosen2004author}, for $n^{th}$ word, we sample a user $w$ uniformly from $\mathbb{U}_s$ and draw the topic indicator variable $z_n^{(s)}$ and word $x_n^{(s)}$ from $G_w$ as follows:
\begin{equation*}
    w \sim \text{Uniform}(\mathbb{U}_s), \quad z^{(s)}_n | G_w \sim G_w, \quad  x^{(s)}_n | z^{(s)}_n \sim \text{Mult.}(\phi_{z^{(s)}_n}). 
\end{equation*}

\xhdr{Generating homogeneity indices of stories}
In equation~\ref{g_u^a}, we assume that a user's probability measure is generated from her preceding user's probability measure and the similarity of the two probability measures is calibrated by the homogeneity index of a news story that bridges the two users.
Here, the homogeneity index of a news story is generated by $\bar{z}^{(s)} = \tfrac{1}{N_s} \sum^{N_s}_{n=1} z^{(s)}_n $, which is a normalization of indicator one-hot vectors in the topic space.
Following the Bayesian Gaussian process latent variable model (Bayesian GP-LVM)~\cite{titsias2010bayesian,damianou2016variational}, we first generate a hidden input vector $c^{(s)}$ from the Gaussian distribution centered at $\bar{z}^{(s)}$, and draw latent function values $\bm{f} \in \mathbb{R}^L$ from the Gaussian distribution centered at zero and the covariance matrix constructed from the pairings of news stories' $c^{(s)}$s. 
Finally, the homogeneity index for a news story $h^{(s)}$ is a noisy observation of $f^{(s)}$ with Gaussian noise. The formal constructions are as follows: 
\begin{equation}
    c^{(s)} \vert \bar{z}^{(s)} \sim \mathcal{N}(\bar{z}^{(s)}, \zeta^{-1} I), \quad \bm{f} \vert \bm{c} \sim \mathcal{N}(0, K_{LL}), \quad \bm{h} | \bm{f} \sim \mathcal{N}(\bm{f}, \kappa^{-1} I).
    \label{homogeneity_value}
\end{equation}
We use squared exponential kernel $k(l, l^\prime) = \sigma^2 \exp{ ( -\tfrac{1}{2}   {\norm{c - c^\prime}}^2 ) }$ for the covariance matrix $K_{LL}$.

\xhdr{Supervised HBTP} 
HBTP can be used for supervised learning, \textit{e.g.}, to predict the type of genuineness of news stories.
Similar to drawing the homogeneity indices, we draw labels of the stories from the same hidden input vector. 
The difference is that the homogeneity indices are hidden in the modeling state and have to be inferred from the model, while the labels are given in the training data.

\xhdr{Parametric counterpart of HBTP}
There is a much overlap in the modeling scheme when we develop a parametric version of HBTP. The major differences are that in the parametric model, (1) we predefine the number of topics and (2) we use the logistic normal prior~\cite{aitchison1982statistical} for the user-topic distributions used to capture the transmissive property among users. With the parametric HBTP, we sacrifice the model flexibility, but we gain an algorithmic benefit of inferring the homogeneity variables for news stories with closed-form updates by solving the Lambert W function. Refer to the Appendix for the generative process of parametric HBTP. 

\xhdr{Remark 2}
In our model, we jointly model Bayesian GP-LVM with a nonparametric topic model which provides both computational and modeling benefits.
First, incorporating GP-LVM with hidden inputs reduces the time complexity of drawing function values $f^{(s)}$ from $\mathcal{O}(L^3)$ to $\mathcal{O}(L P^2)$, where $L$ is the number of stories and $P$ is the number of auxiliary inducing variables~\cite{csato2002sparse,snelson2006sparse,titsias2009variational,seeger2003fast}. 
%
%
In the modeling aspect, we can model multi-dimensional or categorical labels from the topic indicator variable $z^{(s)}$ without directly conditioning on the document-topic simplex, which has been pointed out as one of the weaknesses of the supervised topic model~\cite{mcauliffe2008supervised} compared to the labeled LDA~\cite{ramage2009labeled}.
%
%
Finally, we remark that when computing vector-wise norms of $c^{(s)}$, we resort to a truncated approximation method~\cite{blei2006variational,kurihara2007collapsed,teh2008collapsed} to predefine the upper-bound of the number of topics used. \vspace{-3mm}

\section{Posterior Inference of HBTP}
%
We derive a joint variational inference algorithm to approximate the posterior of HBTP. Since both the text modeling and the homogeneity generation of HBTP fall into the Bayesian nonparametrics, we refer to the well-established literature for variational inference in nonparametric topic models~\cite{blei2006variational,kurihara2007collapsed,teh2008collapsed} and Bayesian GP-LVMs~\cite{damianou2016variational}. We derive the log joint distribution of HBTP as:
\begin{align}
    \ln  p(\bm{x}, \bm{z}, \bm{c}, \bm{h}, \Pi, V, \Phi)  &=  \sum^{L}_{s = 1}  \sum^{N_s}_{n = 1} \Bigg \{ \ln p(x_{n}^{(s)} \vert z_{n}^{(s)}, \Phi) +  \ln \Big ( \frac{1}{\vert \mathbb{U}_s \vert } \sum_{w \in \mathbb{U}_s} p (z_{n}^{(s)} \vert \pi^{(w)} )  \Big ) \Bigg \}  \nonumber \\
    & + \ln p (\bm{h} \vert \bm{c}) + \sum^{L}_{s=1}  \ln p(c^{(s)} \vert z^{(s)}) + \sum^{\infty}_{k=1} \Bigg \{ \ln p (V_k \vert \alpha) + \ln p(\phi_k \vert \alpha_0) \Bigg \} \nonumber \\
    & + \sum^{M}_{u=1} \sum^{\infty}_{k=1} \ln \Big ( \frac{1}{\vert \mathbb{A}_u  \vert} \sum_{v \in \mathbb{A}_u} p(\pi^{(u,v)}_k  \vert V_k, \Pi_k, h^{(u,v)}, \beta)  \Big ) \nonumber.
\end{align}
The log variational distribution of HBTP can be factorized into \vspace{-2mm}
\begin{align}
    \ln q (\bm{z}, \bm{c}, \bm{f}, \Pi, V, \Phi) &= \sum^{L}_{s=1} \sum^{N_s}_{n=1} \ln q(z^{(s)}_n) +  \ln q (\bm{f}) + \sum^{L}_{s=1} \ln q (c^{(s)} ) \nonumber \\ 
     &+ \sum^{M}_{u=1} \sum^{T}_{k=1} \ln \Big ( \frac{1}{\vert \mathbb{A}_u \vert } \sum_{v \in \mathbb{A}_u} q(\pi^{(u,v)}_{k}) \Big ) + \sum^{T}_{k=1} \Bigg \{  \ln q(V_k) + \ln q (\phi_k) \Bigg\} \nonumber,
\end{align}
with the truncation value $T$ with $V_T = 1$. 
We specify the variational distributions for the topic-related latent variables with their variational parameters as \vspace{-1mm}
\begin{align}
    q(z^{(s)}_{n}) &= \text{Multinomial}(z^{(s)}_{n} \vert \gamma^{(s)}_{n}), \nonumber \\ q(\pi^{(u, v)}_k) &= \text{Gamma} (\pi^{(u, v)}_k \vert a^{(u, v)}_k, b^{(u, v)}_k). \nonumber
\end{align}
To specify the variational distributions for homogeneity-related variables, we introduce $G$ auxiliary inducing variables $y \in \mathbb{R}^G$ with corresponding inducing input locations $Y \in \mathbb{R}^{G \times T}$ with $p(y \vert Y) = \mathcal{N}(y \vert 0, K_{GG})$, and express the variational distributions for the hidden input vectors $c^{(s)}$ and noise-free GP latent functions as \vspace{-1mm}
\begin{align*}
    q(c_k^{(s)}) &= \mathcal{N} \Big (c_k^{(s)} \vert \lambda_k^{(s)}, \xi^{-1}_{s,k} I \Big ), \nonumber \\
    q(f^{(s)}) &= \mathcal{N} \Big (f^{(s)} \vert K^\top_{LG} K^{-1}_{GG} y, \quad K_{LL} - K_{LG}K^{-1}_{GG}K_{GL}  \Big ),
\end{align*}
where we set $q(y) = \mathcal{N}(y \vert \mu^{(y)}, \Sigma^{(s)})$ and assume independence among topics for $c^{(s)}$. Here we set $G$ to be drastically smaller than $L$ to increase the inference speed.
Finally, we specify the variational distributions for topic variables as follows:
\begin{equation*}
    q(V_k) = \delta_{V_k}, \quad q(\phi_k) = \text{Dirichlet} (\phi_k \vert \eta_k),
\end{equation*}
where we use the delta function for both simplicity and tractability in inference steps as demonstrated in the work of Liang et al.~\cite{liang2007infinite}. Given the evidence, the task of approximating the variational distribution with respect to the original posterior is equivalent to minimizing the KL divergence between the joint distribution and the factorized variational distribution or maximizing the evidence lower bound. The main derivations come from the expectations of the log-probability of the latent variables with respect to the variational distribution $q$, and the optimization is done using the coordinate ascent algorithm. For the remainder of the section, we report updating formulas for the latent variables that are unique for HTBP compared to previous nonparametric topic models and GP-LVMs. Refer to the Appendix for the full updates. 

\xhdr{Story-level updates} For the story-level latent variables, $z^{(s)}$ is at the intersection of the topic modeling part and the GP part, and the update for $\gamma^{(s)}_{nk}$ is
\begin{equation*}
    \gamma^{(s)}_{n,k} \propto \exp \bigg \{ \mathbb{E}_q[\ln \eta_{k, x^{(s)}_n}] + \mathbb{E}_q[\ln \pi^{(s)}_k] + \frac{\zeta}{N_s} \mathbb{E}_q[c^{(s)}] - \frac{\zeta}{2 N^2_s} \Big ( \sum_{n^\prime \neq n} \gamma^{(s)}_{n^\prime, k} + 1 \Big ) \bigg \}. 
\end{equation*}
Also, the update for the homogeneity variable $h^{(s)}$ is
\begin{equation*}
    \frac{\partial \mathcal{L}}{\partial h^{(s)}} =  \sum^{T}_{k=1} \hspace{-1mm} \sum_{\substack{ u, v :  s^{(u,v)} = s}} \beta  e^{h^{(u,v)}} p^{(v)}_k \bigg \{ \mathbb{E}_q [\ln \pi^{(u,v)}_{k}] - \psi \Big (\beta e^{h^{(u,v)}} p^{(v)}_k \Big ) \bigg \}   
    + \sum^{L}_{s^\prime=1} h^{(s^\prime)} \Big ( W_{s, s^\prime} + W_{s^\prime, s} \Big ),
\end{equation*}
where the full expression for the $L \times L$ matrix $W$ is specified in the Appendix.

\xhdr{User-level updates} Updating $\pi_k^{(u,v)}$ is done as follows:
\begin{equation*}
    a^{(u,v)}_k = \beta \exp(h^{(u,v)}) p^{(v)}_k + \sum^{N_{s^{}}}_{n=1}\gamma^{s^{}}_{n, k}, \quad b^{(u,v)}_k = 1 + \frac{N_s}{\chi_s},
\end{equation*}
where $\chi_s = \sum^{T}_{k=1} \mathbb{E}_q [\pi^{(u,v)}_k]$ and we simplify $s^{(u,v)}$ to $s$.

\xhdr{Computational complexity} For text modeling, HBTP has the time complexity $\mathcal{O}(\vert \mathbb{E} \vert N T + VT)$ with $\vert \mathbb{E} \vert$ being the number of events, $N$ the average length of a story (in words) and $T$ the truncation size. The time complexity for the homogeneity variables is $\mathcal{O}(L P^2)$. \vspace{-3mm}
\section{Experiments}
\begin{figure}[t!]
    \centering
    \includegraphics[width=0.5\linewidth]{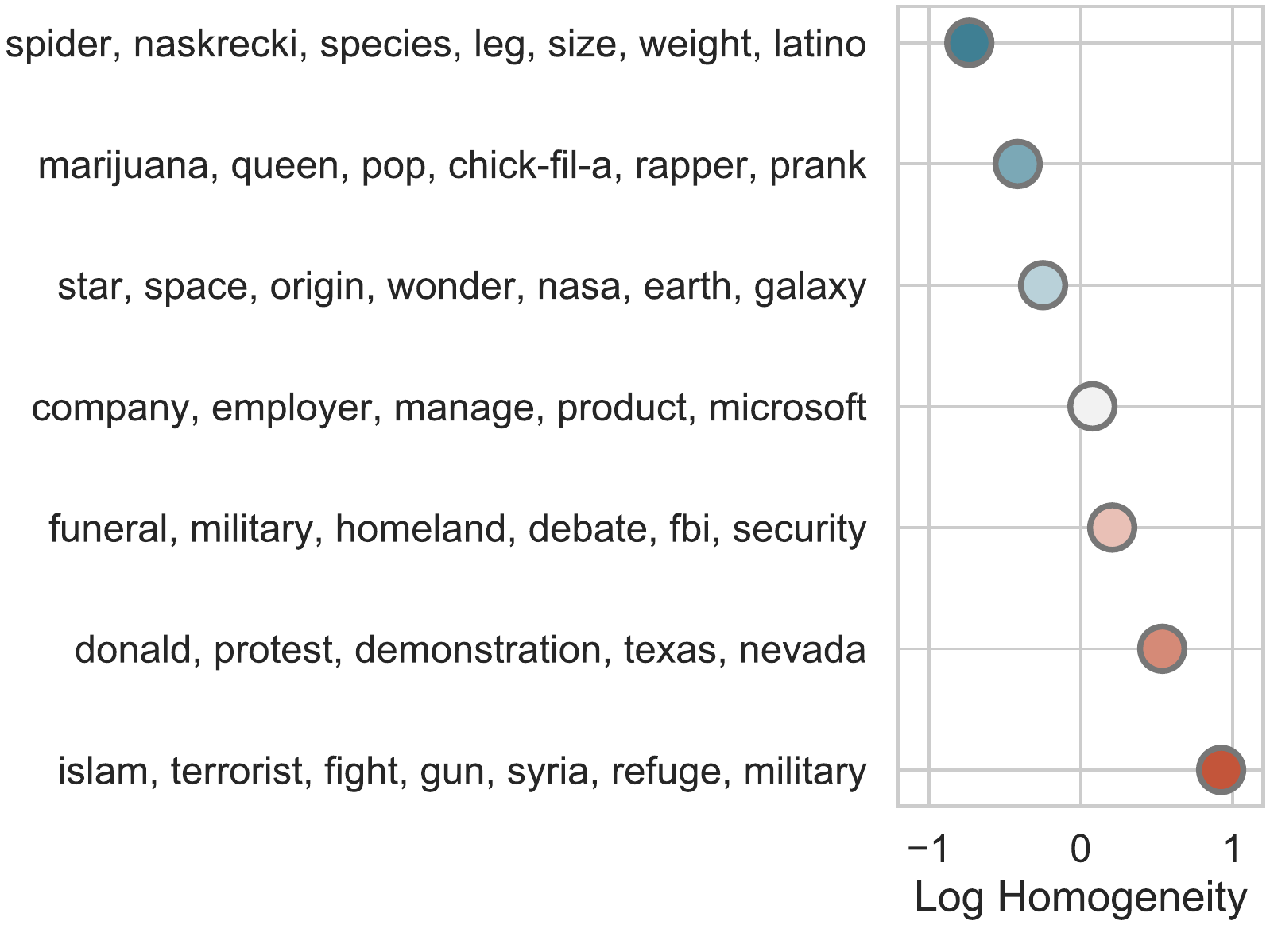} \vspace{-2mm}
    \caption{Topics retrieved by HBTP and their corresponding homogeneity indices. \vspace{-2mm}}
    \label{figure:topic_hval} 
\end{figure}
We first investigate the interplay among news stories' content, homogeneity indices and users' topical interests modeled in nonparametric HBTP and examine how the homogeneity indices of news stories regulate the alignment of topical interests of users who tweet/retweet the news stories. 
Through the process, we find interesting relationship between the homogeneity indices and the labels (genuineness) of news stories when the labels are not observed during training.
Motivated by the finding, we demonstrate how the supervised version of our model (sHBTP) is able to detect the labels of news stories better than the state-of-the-art comparison models.

\xhdr{Dataset} We use the extension of Twitter rumor dataset that has been previously used in rumor detection research~\cite{liu2015real,ma2016detecting,ma2017detect,ma2018rumor}. The original dataset contains, for each tweet, (1) the tweet ID, (2) URL of the news story (3) users' tweet and retweet logs, and (4) a label, which is either true/T, false/F, non-rumor/NR, or unverified/U. 
The "true", "false", and "unverified" stories are potential "rumors" that need to be fact-checked by the debunking websites such as \url{snopes.com} to inspect their genuineness.
"True" stories are the ones that turned out to be genuine after the fact-checking, while "false" stories are the ones that contain misinformation and "unverified" stories are the ones that cannot be conclusively judged even after the human evaluation. 
Finally, "non-rumor" stories are the ones that do not need the fact-checking because \textit{e.g.}, they came from reliable sources.
\begin{figure}[t!]
    \centering
    \begin{subfigure}[b]{0.35\linewidth}
        \centering
        \includegraphics[width=\linewidth]{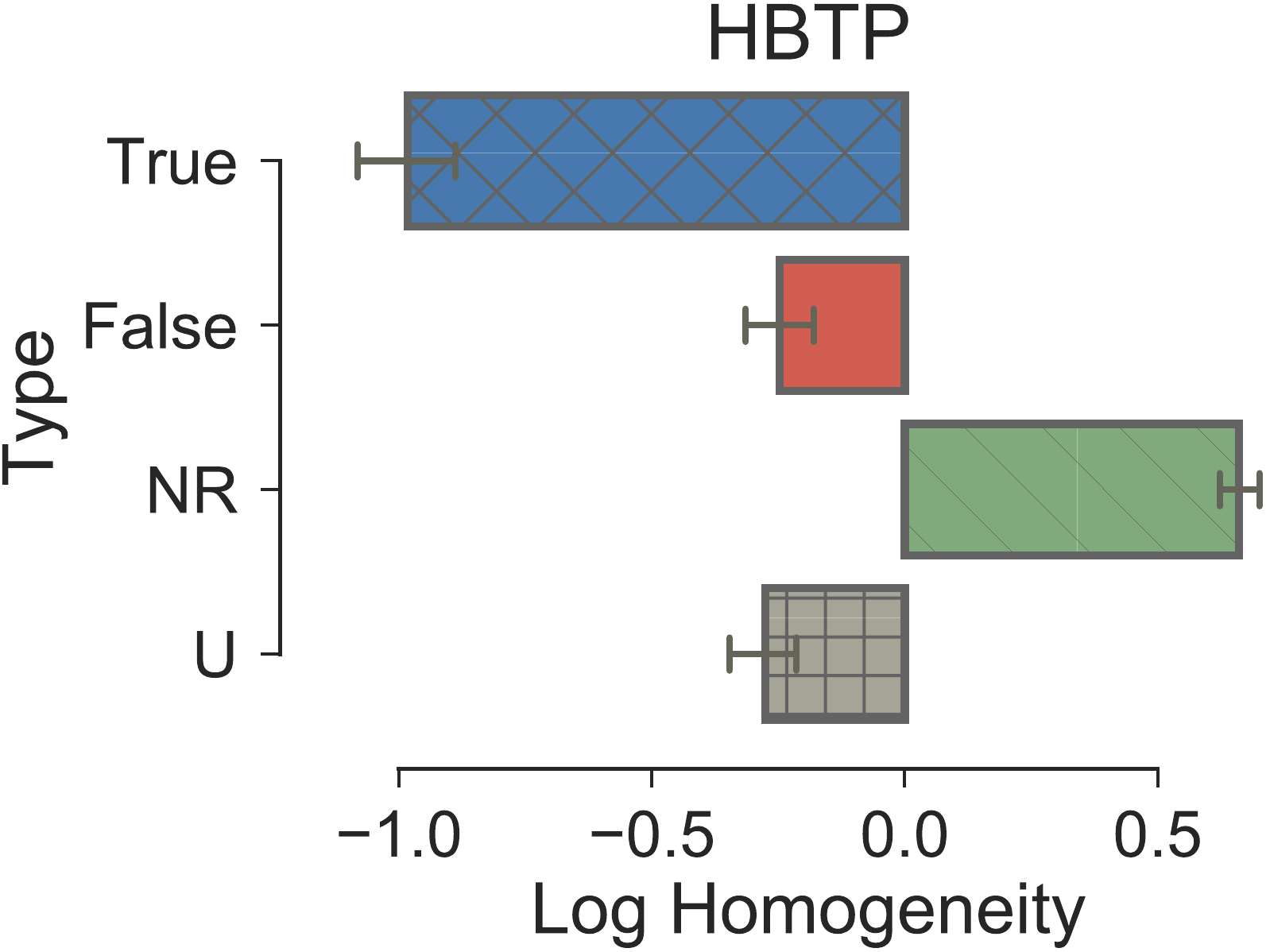} 
        \caption{}
        \label{subfigure:h_index_label}
    \end{subfigure}
    \begin{subfigure}[b]{0.6\linewidth}
        \centering
        \includegraphics[width=\linewidth]{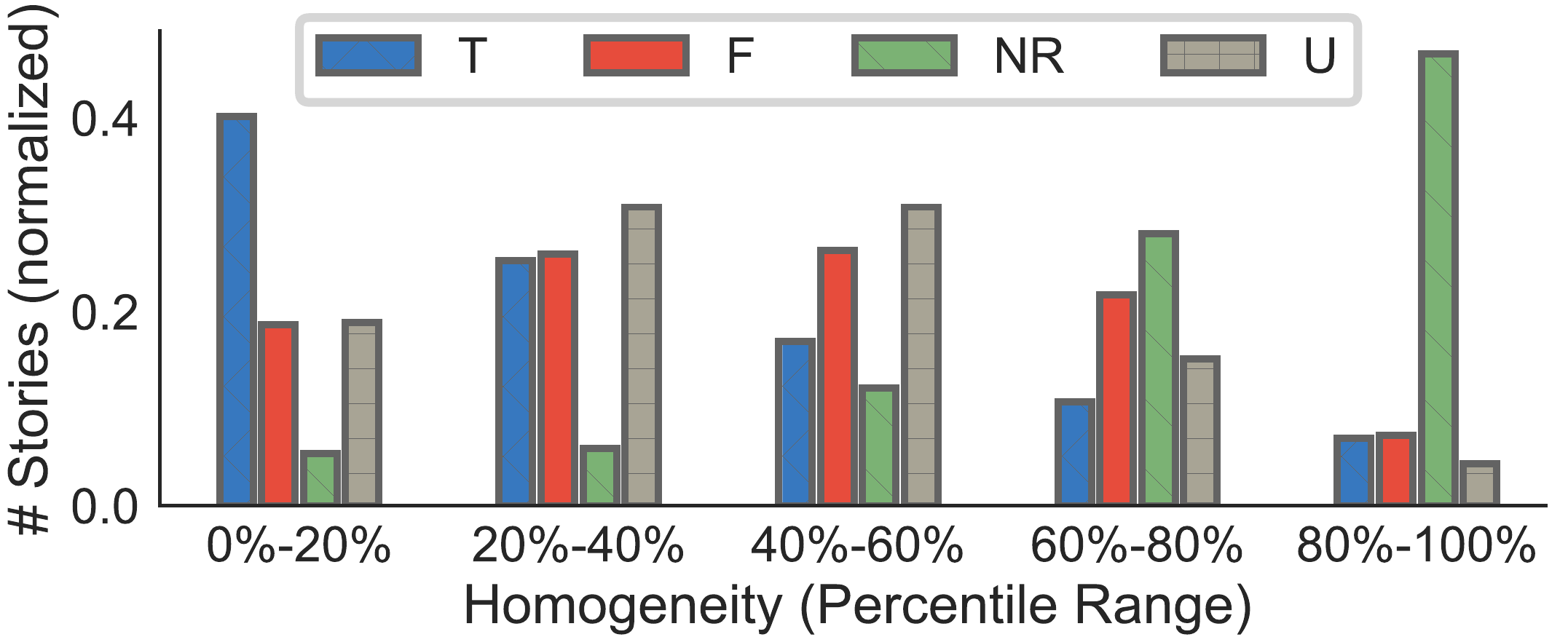}
        \caption{}
        \label{subfigure:figure_hbtp_label}
    \end{subfigure}
    \caption{Relationship between the homogeneity indices by HBTP and the labels (genuineness) of news stories when the labels have not been used during the training. 
    (a) Mean homogeneity indices for stories with different labels retrieved using \textit{unsupervised} HBTP. Homogeneity indices of news stories vary significantly according to their labels expect for "False" and "Unverified" stories.
    (b) Distribution of news stories with different labels over the homogeneity indices.
    \label{figure:figure_hbtp_label}
    \vspace{-2mm}}
\end{figure}
\begin{figure}[t!]
    \centering
    \includegraphics[width=0.55 \linewidth]{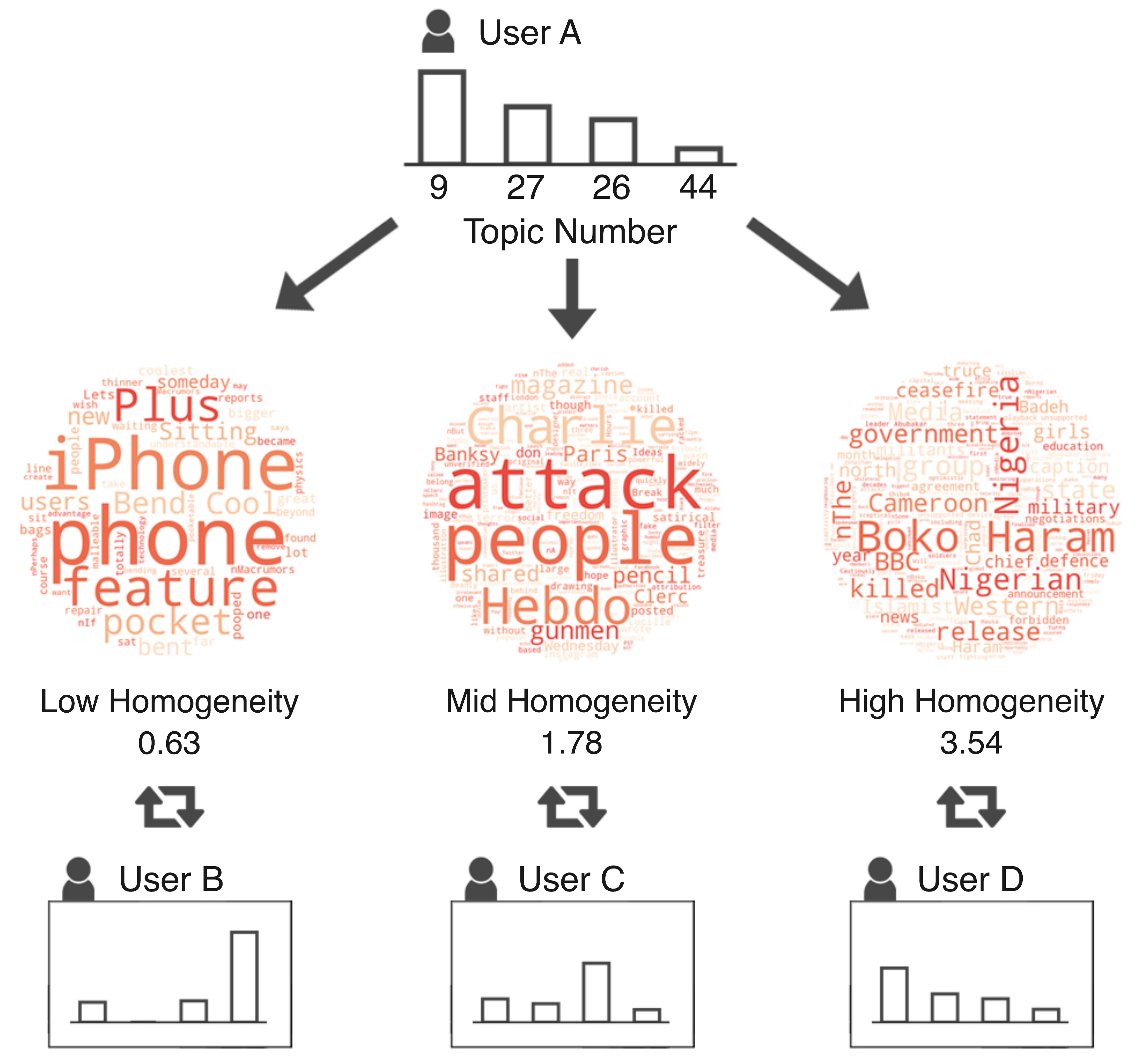}
    \caption{Three example false news stories with different homogeneity indices. All three news stories are originally tweeted by user A, but retweeted by different users B, C, and D respectively. Although all three stories have same labels, their contents are different and have different homogeneity indices, which affect the topical similarities between the tweeting user A and retweeting users B, C, and D. 
    } \vspace{-2mm}
    \label{fig:hbtp_qualitative}
\end{figure}
%
%

%
For the pre-processing of the original dataset, we exclude the news stories whose content information cannot be retrieved with the Twitter API.
Then, we remove "leaf" users, the ones who shared a story once and are at the leaf of the diffusion cascades, since they do not affect the calculation of the homogeneity indices and thus, the outcome of our modeling. 
Also, we confirm that there is no "circular" retweets among users regarding news stories so that our graphical model remains a directed acyclic graph (DAG).
After the pre-processing, there are 79,416 users and 175,389 tweets and retweets. 
For the news stories, there are 1,107 stories in total that contain 12,515 unique tokens. The news stories are divided into 289, 262, 371, and 185 true, false, non-rumor and unverified stories. 
Finally, for the prediction task, we conduct 5-fold cross validation on the Twitter dataset. 

\xhdr{Parameter settings}
For HBTP and sHBTP, we set the topic Dirichlet prior $\alpha_0$ to 0.1. For GP-LVM in the model, we set 50 inducing points for the homogeneity variable and the labels, noise precision parameters $\zeta$, $\kappa$ to 10, and the variational precision parameter $\xi$ to 0.1.
For the comparison methods, we set the values to be consistent with ours for overlapping parameters.
Otherwise, we follow the parameter settings disclosed by the original paper or optimize the values through exhaustive search.

\xhdr{Analyses on the homogeneity of news stories and their topic and labels using unsupervised HBTP}
We analyze the interplay between the homogeneity indices of news stories uncovered by HBTP and the alignment of topical interests of users linked to the story and examine how news stories with different topics and labels have different homogeneity indices.

First, figure~\ref{figure:topic_hval} lists 6 topics chosen from HBTP and their corresponding homogeneity indices.
We observe that topics related to nature, urban culture, and science in general have low homogeneity indices, suggesting that users who share news stories with these topics are in general less topically aligned than those who share other news stories.
On the other hand, the homogeneity indices of economic and industry-related topics are in the middle-range, and topics regarding domestic and international politics have high homogeneity indices.
Note that our topic-word probability measures are attached with homogeneity indices in a similar way to that of supervised topic models~\cite{mcauliffe2008supervised}, since both HBTP and supervised topic models leverage the labels conditioned on topics, unlike other models that incorporate label information but the conditioning is the other way around~\cite{ramage2009labeled}.
However, the modeling assumption of HBTP diverges from that of supervised topic models in that in supervised topic models, the labels are "given",
whereas in our model, both topics and homogeneity indices are "latent", and the homogeneity indices need to be inferred from users' tweet/retweeting patterns and both topics and homogeneity indices mutually reshape one another during the joint-modeling.
Second, from figure~\ref{subfigure:h_index_label}, we observe that the homogeneity indices of stories vary significantly according to their labels.
We note that HBTP computes the homogeneity indices of news stories in unsupervised fashion, without observing the labels of the stories while training.
Interestingly, "non-rumors" have the highest homogeneity indices while the other types of news stories that went through fact-checking process have much lower values. 
Among the fact-checked stories, the stories that are verified to be "true" have significantly lower homogeneity indices than "false" and "unverified" stories.
Figure~\ref{subfigure:figure_hbtp_label} shows the distribution of news stories with different labels over homogeneity indices.
From the figure, we confirm that majority of "non-rumors" have high homogeneity indices (top 40\%), majority of "true" stories have low homogeneity indices (bottom 40\%), while "false" and "unverified" stories are distributed in the middle range.
Overall, the difference in homogeneity indices of news stories with respect to different labels suggest that we can leverage such homogeneity indices and topics to classify the news labels in prediction tasks.
%
%
\begin{table*}[t!]
\centering
    \caption{Results on label classification task using supervised HBTP and other methods. The plus-minus sign indicates one standard error.} \vspace{-2mm}
    \begin{tabular}{@{}lrrrrr @{}} \toprule
    \multirow{2}{*}{Method} & \multicolumn{1}{l}{\multirow{2}{*}{Accuracy}} & \multicolumn{1}{r}{\quad\quad\quad True}    & \multicolumn{1}{r}{\quad\quad\quad False}    & \multicolumn{1}{r}{Non-Rumor}   & \multicolumn{1}{r}{Unverified}    \\
                            & \multicolumn{1}{l}{}                      & \multicolumn{1}{r}{$F_1$} & \multicolumn{1}{r}{$F_1$} & \multicolumn{1}{r}{$F_1$} & \multicolumn{1}{r}{$F_1$} \\ \midrule
    HDP+SVM: Linear              & 0.484 $\pm$ 0.019                                     & 0.593                    & 0.490                    & 0.514                    & 0.357\\
    HDP+SVM: RBF              & 0.605 $\pm$ 0.024                                     & 0.644                    & 0.522                    & 0.629                    & 0.514\vspace{2mm}\\ 
    HBTP+SVM: Linear                   & 0.533 $\pm$ 0.014                                     & 0.621                    & 0.464                    & 0.641                    & 0.593                   \\ 
    HBTP+SVM: RBF                   & 0.717 $\pm$ 0.006                                     & 0.855                    & 0.550                    & 0.682                    & 0.653\vspace{2mm} \\ 
    GPSTM (\small{Kandemir \textit{et al.}, 2018})                 & 0.664 $\pm$ 0.013                                     & 0.662                    & 0.686                    & 0.687                    & 0.435                      \\
    BU-RvNN (\small{Ma \textit{et al.}, 2018})                 & 0.622 $\pm$ 0.010                                     & 0.616                    & 0.571                    & 0.687                    & 0.584                      \\
    TD-RvNN (\small{Ma \textit{et al.}, 2018})                 & 0.698 $\pm$ 0.004                                     & 0.664                    & 0.668                    & 0.723                    & \textbf{0.693}\vspace{2mm} \\
    Supervised HBTP                   & \textbf{0.781 $\pm$ 0.012}                                 & \textbf{0.891}                   & \textbf{0.740}                    & \textbf{0.812}                    & 0.622                   \\ \bottomrule
    \end{tabular} \vspace{-3mm}
  \label{table:shbtp}
\end{table*}

Lastly, we highlight the \textit{transmissive} nature of our model and inspect how the homogeneity of a news story affects the degree of transmission between the topical interests of user pairs who tweet and retweet the story. 
In figure~\ref{fig:hbtp_qualitative}, we visualize three news stories tweeted by a user (user A).
All three stories are labeled as "false", but the content of the stories differ, leading to different homogeneity indices.
Here, the topic proportion of user A is transmitted to three other users, B, C, and D, each of whom retweeted different tweets of user A with different stories.
Among users B, C, and D, since user D retweeted a tweet that contains a high homogeneity-valued story, her topic distribution is more similar to user A than, say, that of users B and C. 
By generalizing this observation to multiple tweeting and retweeting users, we can presume that the topical interests of users that share a high homogeneity-valued story will be more uniform than that of the users who share news stories with low homogeneity indices.

\vspace{2mm}
\xhdr{Story label prediction using supervised HBTP}
We validate our modeling scheme by showing the predictive power of our model compared with the other competitive models.
For this classification task, we use sHBTP that observes the labels of the news stories in the training set.
Our comparison models are as follows:
\begin{itemize}
    \item \textbf{HDP + SVM}: 
    This method uses document-topic probability measures for news stories using HDP, and uses support vector machine (SVM) as a classifier. 
    For SVM, we use linear and radial basis function (RBF) kernels.
    \item \textbf{HBTP + SVM}: 
    Unlike HDP, document-topic probability measures are drawn per users, and the topics of news stories are represented by aggregating users' probability measures who shared the story. 
    Also, the homogeneity of news stories are incorporated in the model which affect users' probability measures.
    \item \textbf{GPSTM}:
    Gaussian process supervised topic models~\cite{kandemir2017supervising} extend parametric supervised topic models~\cite{mcauliffe2008supervised} by using Bayesian GP-LVMs as classifiers instead of linear classifiers and the labels are conditioned on news stories' topics.
    \item \textbf{BU-RvNN} and \textbf{TD-RvNN}: 
    Bottom-up and Top-down recursive neural networks~\cite{ma2018rumor} are the state-of-the-art models designed for classifying true and false news. 
    The models are operationalized on bottom-up and top-down propagation trees where each node contains tweet contents.
    Thus, our model and the other topic-based comparison models exploit the content information of news sources, whereas BU-RvNN and TD-RvNN use content information of tweets.
\end{itemize}
Note that there are trade-offs between using the content information of news stories and tweets.
\begin{figure*}[t!]
    \centering
    \begin{subfigure}[b]{0.21\linewidth}
        \centering
        \includegraphics[width=\linewidth]{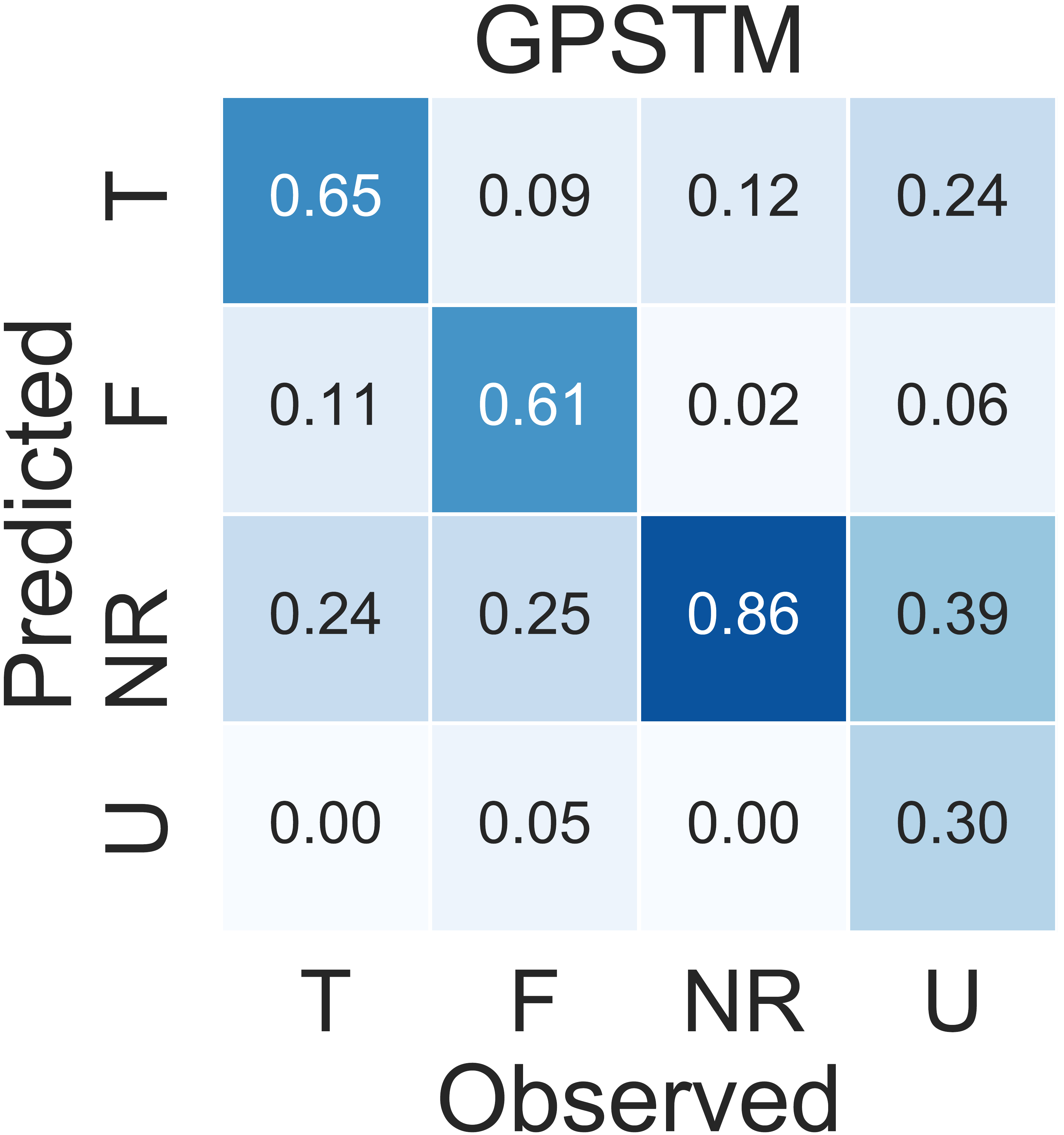}
        \label{fig:confusion_svmrbf}
    \end{subfigure}
    \hfill
    \begin{subfigure}[b]{0.21\linewidth}
        \centering
        \includegraphics[width=\linewidth]{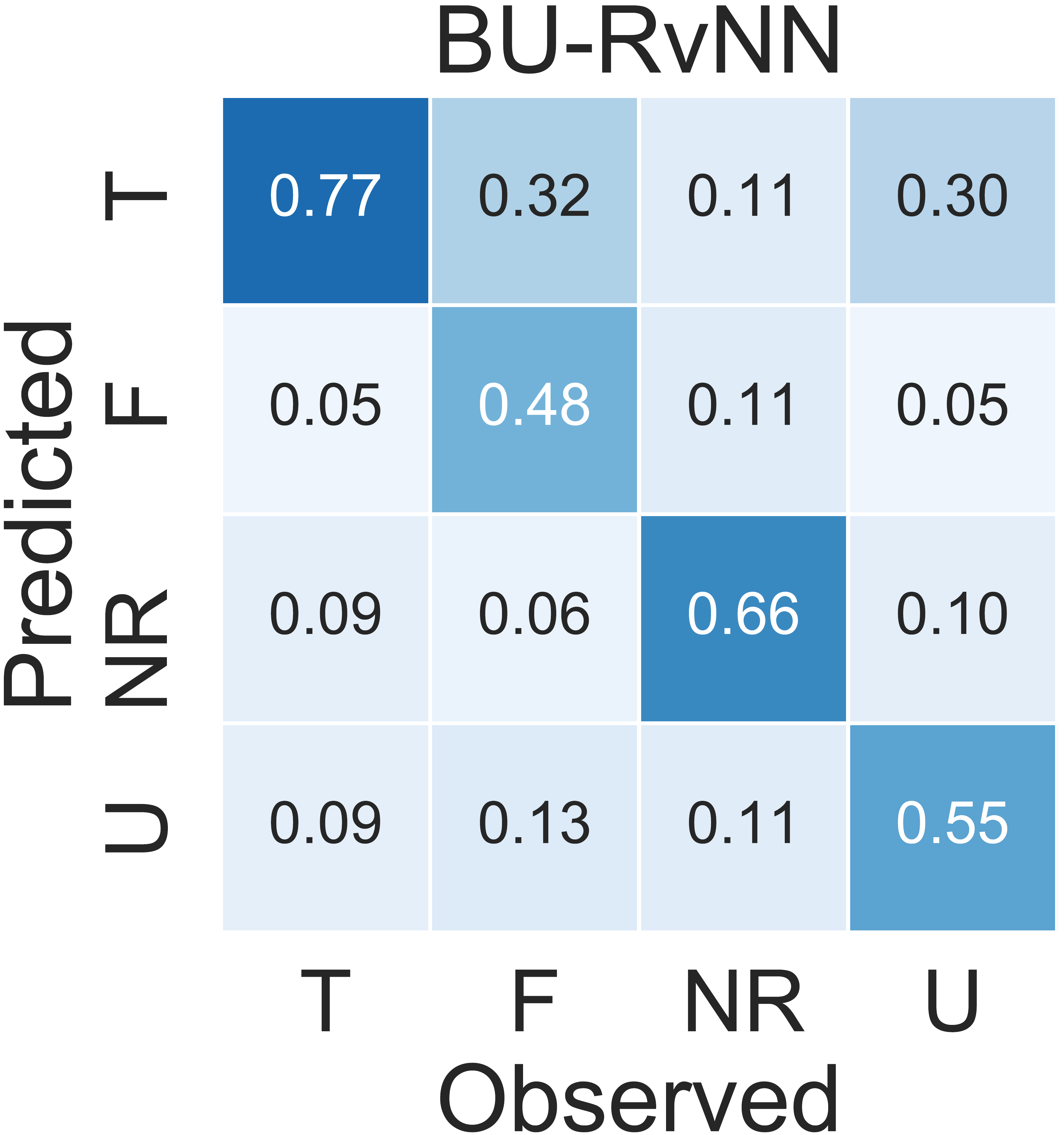}
        \label{fig:confusion_burvnn}
    \end{subfigure}
    \hfill
    \begin{subfigure}[b]{0.21\linewidth}
        \centering
        \includegraphics[width=\linewidth]{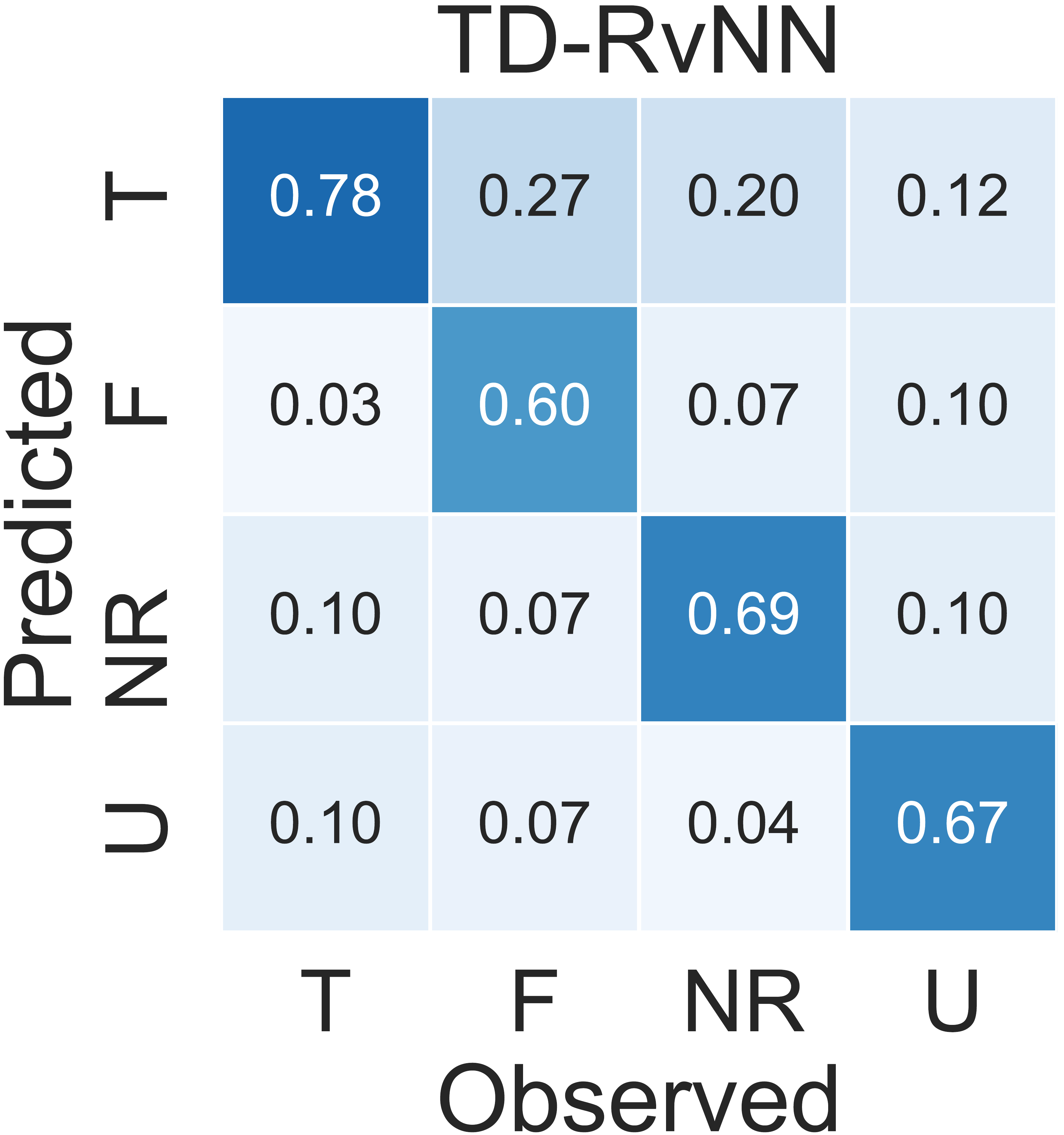}
        \label{fig:confusion_tdrvnn}
    \end{subfigure}
    \hfill
    \begin{subfigure}[b]{0.21\linewidth}
        \centering
        \includegraphics[width=\linewidth]{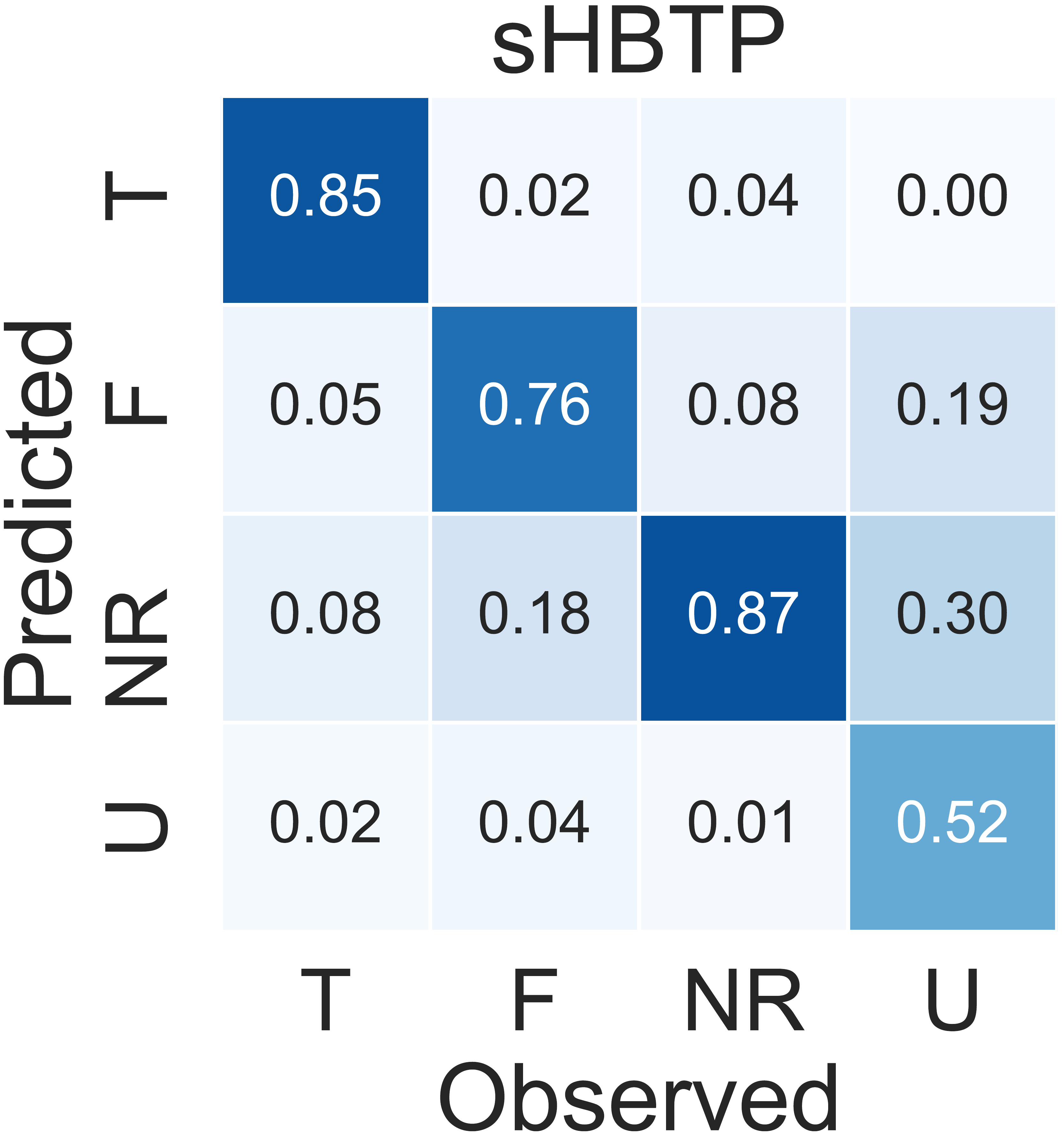}
        \label{fig:confusion_shbtp}
    \end{subfigure}
    \vspace{-5mm}
    \caption{Confusion matrices for classification results of GPSTM, BU-RvNN, TD-RvNN, and supervised HBTP.
    }
    \label{fig:confusion}
\end{figure*}
From one standpoint, contents in news stories have richer description about the actual event from which the models judge the genuineness.
From another standpoint, as stated in the work of Ma \textit{et al.}~\cite{ma2017detect}, the tweets often contain user judgments or debates regarding the news stories' genuineness that can be critical hints the models exploit.
%
%

%
In table~\ref{table:shbtp}, we report the overall classification accuracy and the $F_1$ scores for the four labels.
We confirm that our model outperforms the comparison methods in the overall accuracy and in the $F_1$ scores for all labels except "unverified". 
Specifically, in figure~\ref{fig:confusion}, we confirm that our model excels at differentiating "true" and "false" news, two controversial groups of news stories that needed to be fact-checked by human experts for the classification. 
On the flip side, our model is relatively poor at detecting "unverified" stories.
One possible reason for the performance edge of our model is the inclusion of homogeneity variable in our model;
as observed in figure~\ref{figure:figure_hbtp_label}, the log homogeneity indices of stories with four labels differ significantly, which help the model to classify confusing label pairs, \textit{e.g.}, high-valued non-diagonal entries in the confusion matrices in figure~\ref{fig:confusion}. 
Also, the performance edge over GPSTM highlights the efficacy of our model's nonparametric and joint modeling-nature among user interests, news stories' contents and the homogeneity variables that encode information of users' sharing patterns. \vspace{-2mm}
\section{Conclusion and Future Work}
We developed HBTP, a novel Bayesian nonparametric model that jointly models topics and homogeneity indices of news stories and user interests. HBTP for modeling the content of news stories extends hierarchical Dirichlet process in a way that instead of a single upper-level probability measure affecting the lower-level probability measures, we allowed the probability measures of the bottom-layer to be transmitted from one another. To discover homogeneity values of the news stories, we incorporated Bayesian GP-LVM in the model. By conducting quantitative and qualitative analyses, we found interesting relationships between the homogeneity index and the genuineness of news stories, and between the homogeneity index and the topics extracted. Finally, we showed how HBTP can be extended easily to predict genuineness of news stories better than the state-of-the-art rumor detection methods. 

Our model can be extended to leverage different types of data to increase the flexibility of the model and to boost the predictive power.
One direction is to explicitly incorporate users' diffusion cascades and the contents of the user tweets in the model. 
For example, our model and RvNN based model of Ma \textit{et al.}~\cite{ma2018rumor} can work in a complementary fashion because the input spaces these models operate on are almost orthogonal to each other.

Also, while the RvNN based models require the users' diffusion cascades and thus cannot predict completely new news stories, we can explore the predictive power of our model on newly created stories with no user sharing by directly inferring the topic probability measures of these held-out stories. 
In this setting, we can only leverage the homogeneity indices of the training stories. \vspace{-3mm}
%


\begin{appendices}
\section{Parametric HBTP}
For parametric HBTP, we sample a user's topic distribution $\theta$ using a multivariate normal distribution with a mean being the preceding user's topic distribution. For the variance, we incorporate the homogeneity value to regulate the similarity between the two users' topic distributions.
Then, we put a sigmoid function over $\theta$ to turn it into the simplex, and draw topic labels for each words for news stories.
%
Note that the logistic normal distribution has been previously used in correlated topic models~\cite{blei2005correlated}, but with a different purpose: to capture correlations among topic.
Finally, we illustrate the generative process of the parametric counterpart of HBTP as follows: 
\begin{enumerate}
    \item For each topic $k$, draw word-topic distribution $\phi_k \sim \text{Dir}(\alpha)$.
    \item For user $u$, draw $\theta_u$ with\\$ p(\theta_u) = \tfrac{1}{\vert \mathbb{A}_u \vert} \sum_{v \in \mathbb{A}_u} \mathcal{N}(\theta_u \vert \theta_v, \exp{(- h^{(u, v)})}\Sigma )$ if $u$ has preceding users $v$. If not, draw $ \theta_u \sim \mathcal{N}(\mu, \Sigma )$.
    \item For news story $s$:
        \begin{itemize}
            \item[-] For each word index $n$:
            \begin{itemize}
                \item[] Draw user $w$ who propagated the story $s$.
                \item[] Draw topic indicator variable $z_n^{(s)} \sim \text{Sigmoid}(\theta_w)$ and draw word $x_n^{(s)} \sim \phi_{z_n^{(s)}}$.
            \end{itemize}
            \item[-] Draw homogeneity value $h^{(s)}$ using equation (9).  
        \end{itemize}
\end{enumerate}
\section{Posterior Inference}
\xhdr{Expectations w.r.t. $q$} By taking expectations of the latent variables in HBTP with respect to the variational distributioin $q$, we get the following results: 
\begin{equation*}
    \mathbb{E}_q [\ln p(x_n^{(s)} \vert z_n^{(s)})] = \sum^{T}_{k=1} \gamma^{(s)}_{nk} \mathbb{E}_q [\ln p (x^{(s)}_n \vert \phi_k)] = \sum^{T}_{k=1} \gamma^{(s)}_{nk} \bigg \{   \psi(\eta_{k x^{(s)}_n}) - \psi \Big (\sum^{\vert V \vert}_{i=1}  \eta_{k i} \Big ) \bigg \}.
\end{equation*}
For stories and user variables,
\begin{align*}
    \mathbb{E}_q [\ln p(z_n^{(s)} \vert \pi^{(w)} ) ] &= \sum^{T}_{k=1} \gamma^{(s)}_{nk} \mathbb{E}_q \Big [ \ln p (z^{(s)}_n = k \vert \pi^{(w)})  \Big ] \\
    &= \sum^{T}_{k=1} \gamma^{(s)}_{nk} \bigg \{ \mathbb{E}_q [\ln \pi^{(w)}_k] - \mathbb{E}_q [\ln \sum^{T}_{k^\prime =1} \pi^{(w)}_{k^\prime}]  \bigg \} \\
    &\geq \sum^{T}_{k=1} \gamma^{(s)}_{nk} \bigg \{ \mathbb{E}_q [\ln \pi^{(w)}_k] - \ln \chi - \frac{\sum^{T}_{k=1} \mathbb{E}_q [\pi^{(w)}_k] - \chi}{\chi}  \bigg \} \\
    &\geq \sum^{T}_{k=1} \gamma^{(s)}_{nk} \bigg \{ \mathbb{E}_q [ \psi(a^{(w)}_k) - \ln b^{(w)}_k - \ln \chi - \frac{\sum^{T}_{k=1} \mathbb{E}_q [\pi^{(w)}_k] - \chi}{\chi}  \bigg \},\\
\end{align*} 
\begin{equation*}
    \mathbb{E}_q [\ln p (\pi^{(u,v)}_{k} \vert V_k, \Pi_k, h^{(u,v)}, \beta )] =\Big (\beta e^{h^{(u,v)}} p^{(v)}_k - 1  \Big ) \Big \{ \psi(a^{(u,v)}_k) - \ln b^{(u,v)}_k \Big \} - \frac{a_k^{(u,v)}}{b_k^{(u,v)}} - \ln \Gamma(\beta e^{h^{(u,v)}} p^{(v)}_k).
\end{equation*}
For the corpus-level, 
\begin{equation*}
    \mathbb{E}_q [\ln p (V_k \vert \alpha)] = \sum^{T}_{k=1} \ln \Gamma (\alpha + 1) - \ln \Gamma(\alpha) + (\alpha - 1) \ln (1 - V_k).
\end{equation*}
Finally, for GPLVM latent variables, 
\begin{equation*}
    \mathbb{E}_q [\ln p (c^{(s)} \vert z^{(s)})] = \frac{T}{2} \ln 2 \pi \zeta - \frac{1}{2} \zeta \bigg \{ \sum^{T}_{k=1} \xi^{-1}_{s,k}  + (\lambda^{(s)} - \frac{1}{N_s} \sum^{N_s}_{n=1}\gamma^{(s)}_{n} )^\top (\lambda^{(s)} - \frac{1}{N_s} \sum^{N_s}_{n=1}\gamma^{(s)}_{n} )  \bigg \},
\end{equation*}
\begin{equation*}
    \mathbb{E}_q [\ln p (\bm{h} \vert \bm{c})] \geq \ln \Bigg [ \frac{\kappa^{\frac{L}{2}}{\vert K_{GG} \vert}^{\frac{1}{2}}  }{ (2\pi)^{\frac{L}{2}}{\vert \kappa \psi_2 + K_{GG} \vert}^{\frac{1}{2}} }   \Bigg ] - \frac{1}{2}\bm{h}^\top W \bm{h} - \frac{\kappa}{2}\psi_0 + \frac{\kappa}{2} \text{tr} (K^{-1}_{GG} \psi_2).
\end{equation*}
\xhdr{Corpus-level updates} Updating $\eta_k$ can be done in a closed-form:
\begin{equation*}
    \eta_{k,i} = \alpha_0 + \sum^{L}_{s=1} \sum^{N_s}_{n=1} \gamma^{(s)}_{n,k} \mathbb{I} (x^{(s)}_n = i).
\end{equation*}
To update $V_k$, we use the steepest ascent algorithm:
\begin{equation*}
    \frac{\partial \mathcal{L}}{ \partial V_k } = - \frac{\alpha - 1}{1 - V_k} + \beta \bigg [  \sum^{L}_{s=1} \Big ( \mathbb{E}_q [ \ln \pi^{(s)}_k] - \mathbb{E}_q [\pi^{(s)}_k ] - \psi(\beta p_k)  \Big )  \bigg ] \bigg [  \frac{p_k}{V_k} - \sum_{k^\prime > k} \frac{p_{k^\prime}}{1 - V_k} \bigg ].
\end{equation*}
\xhdr{GPLVM updates} Updating formula for $W$, with $\psi_0 = L \sigma^2$ and $\psi_2 = \sum^{L}_{s=1} \psi_2^{(s)}$, is:
\begin{equation*}
    W = \kappa I - \kappa^2 \psi_1 (\kappa \psi_2 + K_{GG})^{-1} \psi_1^{\top},
\end{equation*}
where
\begin{equation*}
    (\psi_1)_{s, g} = \sigma^2 \prod^{T}_{k=1} \exp \bigg \{ - \frac{1}{2} (\lambda^{(s)}_k - Y^{(g)}_{k})^2 (\xi^{-1}_{s, k} + 1)^{-1}  \bigg \} (\xi^{-1}_{s, k} + 1)^{-\frac{1}{2}},
\end{equation*}
\begin{equation*}
    (\psi_2^{(s)})_{g, g^\prime} = \sigma^4 \prod^{T}_{k=1} \exp \bigg \{ - \frac{1}{4} (Y^{(g)}_{k} - Y^{(g^\prime)}_{k})^2 - \frac{(\lambda^{(s)}_k - \bar{Y}_k )^2}{(2\xi^{-1}_{s, k} + 1) }  \bigg \} (2 \xi^{-1}_{s, k} + 1)^{-\frac{1}{2}}.
\end{equation*}
To update $\mu^{(y)}$ and $\Sigma^{(y)}$, we calculate
\begin{align}
    \mu^{(y)} &= \kappa \Sigma^{(y)} K_{GG}^{-1} \mathbb{E}_q [K_{GL}] \bm{h}, \nonumber \\
    \Sigma^{(y)} &= \Big (K^{-1}_{GG} + \kappa K^{-1}_{GG} \mathbb{E}_q [K_{GL} K_{GL}^\top] K_{GG}^{-1} \Big )^{-1}. \nonumber
\end{align}
Finally, for $\lambda^{(s)}$, we have:
\begin{align}
    &\frac{\partial \mathcal{L}}{\partial \lambda^{(s)}} = -\zeta \lambda^{(s)} + \frac{\zeta}{N_s}\sum^{N_s}_{n=1} \gamma^{(s)}_n + \kappa \sum^{L}_{s=1} h^{(s)} \frac{ \partial E_q[ k_{Gc^{(s)}}]}{\partial \lambda^{(s)}} K_{GG}^{-1} \mu^{(y)} \nonumber \\
    & - \frac{\kappa}{2} \sum^{L}_{s=1} \text{tr} \Big ( \frac{ \partial \mathbb{E}_q [k_{Gc^{(s)}} k^{\top}_{Gc^{(s)}}] + I }{\partial \lambda^{(s)}}   K_{GG}^{-1} \big( \mu^{(y)} {\mu^{(y)}}^\top + \Sigma^{(y)} \big) K_{GG}^{-1} \Big ). \nonumber
\end{align}

\end{appendices}

\clearpage

\bibliographystyle{abbrv}
\bibliography{ref}

\begin{thebibliography}{10}

\bibitem{ahmed2010timeline}
A.~Ahmed and E.~Xing.
\newblock Timeline: A dynamic hierarchical {D}irichlet process model for
  recovering birth/death and evolution of topics in text stream.
\newblock In {\em UAI}, 2010.

\bibitem{aitchison1982statistical}
J.~Aitchison.
\newblock The statistical analysis of compositional data.
\newblock {\em Journal of the Royal Statistical Society. Series B},
  44(2):139--177, 1982.

\bibitem{babaei2018analysing}
M.~Babaei, A.~Chakraborty, J.~Kulshrestha, E.~Redmiles, M.~Cha, and K.~Gummadi.
\newblock Analysing biases in perception of truth in news stories and their
  implications for fact checking.
\newblock 2018.

\bibitem{balmau2018limiting}
O.~Balmau, R.~Guerraoui, A.-M. Kermarrec, A.~Maurer, M.~Pavlovic, and
  W.~Zwaenepoel.
\newblock Limiting the spread of fake news on social media platforms by
  evaluating users' trustworthiness.
\newblock {\em arXiv preprint arXiv:1808.09922}, 2018.

\bibitem{bessi2015science}
A.~Bessi, M.~Coletto, G.~A. Davidescu, A.~Scala, G.~Caldarelli, and
  W.~Quattrociocchi.
\newblock Science vs conspiracy: Collective narratives in the age of
  misinformation.
\newblock {\em PloS one}, 10(2):e0118093, 2015.

\bibitem{bessi2014economy}
A.~Bessi, A.~Scala, L.~Rossi, Q.~Zhang, and W.~Quattrociocchi.
\newblock The economy of attention in the age of (mis) information.
\newblock {\em Journal of Trust Management}, 1(1):12, 2014.

\bibitem{blei2006variational}
D.~Blei and M.~Jordan.
\newblock Variational inference for {D}irichlet process mixtures.
\newblock {\em Bayesian analysis}, 1(1):121--143, 2006.

\bibitem{blei2005correlated}
D.~Blei and J.~Lafferty.
\newblock Correlated topic models.
\newblock In {\em NIPS}, 2005.

\bibitem{budak2011limiting}
C.~Budak, D.~Agrawal, and A.~El~Abbadi.
\newblock Limiting the spread of misinformation in social networks.
\newblock In {\em WWW}, 2011.

\bibitem{csato2002sparse}
L.~Csat{\'o} and M.~Opper.
\newblock Sparse on-line {G}aussian processes.
\newblock {\em Neural computation}, 14(3):641--668, 2002.

\bibitem{damianou2016variational}
A.~Damianou, M.~Titsias, and N.~Lawrence.
\newblock Variational inference for latent variables and uncertain inputs in
  {G}aussian processes.
\newblock {\em JMLR}, 17(1):1425--1486, 2016.

\bibitem{del2016spreading}
M.~Del~Vicario, A.~Bessi, F.~Zollo, F.~Petroni, A.~Scala, G.~Caldarelli,
  E.~Stanley, and W.~Quattrociocchi.
\newblock The spreading of misinformation online.
\newblock {\em PNAS}, 113(3):554--559, 2016.

\bibitem{devroye1986sample}
L.~Devroye.
\newblock Sample-based non-uniform random variate generation.
\newblock In {\em Winter Simulation Conference}, 1986.

\bibitem{dietz2007unsupervised}
L.~Dietz, S.~Bickel, and T.~Scheffer.
\newblock Unsupervised prediction of citation influences.
\newblock In {\em ICML}, 2007.

\bibitem{du2015dirichlet}
N.~Du, M.~Farajtabar, A.~Ahmed, A.~Smola, and L.~Song.
\newblock Dirichlet-{H}awkes processes with applications to clustering
  continuous-time document streams.
\newblock In {\em KDD}, 2015.

\bibitem{du2013uncover}
N.~Du, L.~Song, H.~Woo, and H.~Zha.
\newblock Uncover topic-sensitive information diffusion networks.
\newblock In {\em AISTATS}, 2013.

\bibitem{farajtabar2017fake}
M.~Farajtabar, J.~Yang, X.~Ye, H.~Xu, R.~Trivedi, E.~Khalil, S.~Li, L.~Song,
  and H.~Zha.
\newblock Fake news mitigation via point process based intervention.
\newblock In {\em ICML}, 2017.

\bibitem{ferguson1973bayesian}
T.~Ferguson.
\newblock A {B}ayesian analysis of some nonparametric problems.
\newblock {\em The annals of statistics}, pages 209--230, 1973.

\bibitem{friggeri2014rumor}
A.~Friggeri, L.~Adamic, D.~Eckles, and J.~Cheng.
\newblock Rumor cascades.
\newblock In {\em ICWSM}, 2014.

\bibitem{gupta2013faking}
A.~Gupta, H.~Lamba, P.~Kumaraguru, and A.~Joshi.
\newblock Faking sandy: characterizing and identifying fake images on twitter
  during hurricane sandy.
\newblock In {\em WWW}, 2013.

\bibitem{he2015hawkestopic}
X.~He, T.~Rekatsinas, J.~Foulds, L.~Getoor, and Y.~Liu.
\newblock Hawkestopic: A joint model for network inference and topic modeling
  from text-based cascades.
\newblock In {\em ICML}, 2015.

\bibitem{holcomb2013news}
J.~Holcomb, J.~Gottfried, A.~Mitchell, and J.~Schillinger.
\newblock News use across social media platforms.
\newblock {\em Pew Research Journalism Project}, 2013.

\bibitem{jordan1999introduction}
M.~Jordan, Z.~Ghahramani, T.~Jaakkola, and L.~Saul.
\newblock An introduction to variational methods for graphical models.
\newblock {\em Machine learning}, 37(2):183--233, 1999.

\bibitem{kandemir2017supervising}
M.~Kandemir, T.~Keke{\c{c}}, and R.~Yeniterzi.
\newblock Supervising topic models with {G}aussian processes.
\newblock {\em Pattern Recognition}, 77:226--236, 2018.

\bibitem{kim2017hierarchical}
D.~Kim and A.~Oh.
\newblock Hierarchical {D}irichlet scaling process.
\newblock {\em Machine Learning}, 3(106):387--418, 2017.

\bibitem{kim2017joint}
J.~Kim, D.~Kim, and A.~Oh.
\newblock Joint modeling of topics, citations, and topical authority in
  academic corpora.
\newblock {\em TACL}, 5:191--204, 2017.

\bibitem{kim2018leveraging}
J.~Kim, B.~Tabibian, A.~Oh, B.~Sch{\"o}lkopf, and M.~Gomez~Rodriguez.
\newblock Leveraging the crowd to detect and reduce the spread of fake news and
  misinformation.
\newblock In {\em WSDM}, 2018.

\bibitem{kumar2018false}
S.~Kumar and N.~Shah.
\newblock False information on web and social media: A survey.
\newblock {\em arXiv preprint arXiv:1804.08559}, 2018.

\bibitem{kumar2016disinformation}
S.~Kumar, R.~West, and J.~Leskovec.
\newblock Disinformation on the web: Impact, characteristics, and detection of
  wikipedia hoaxes.
\newblock In {\em WWW}, 2016.

\bibitem{kurihara2007collapsed}
K.~Kurihara, M.~Welling, and Y.~W. Teh.
\newblock Collapsed variational {D}irichlet process mixture models.
\newblock In {\em IJCAI}, 2007.

\bibitem{kwon2017rumor}
S.~Kwon, M.~Cha, and K.~Jung.
\newblock Rumor detection over varying time windows.
\newblock {\em PloS one}, 12(1):e0168344, 2017.

\bibitem{kwon2013prominent}
S.~Kwon, M.~Cha, K.~Jung, W.~Chen, et~al.
\newblock Prominent features of rumor propagation in online social media.
\newblock In {\em ICDM}, 2013.

\bibitem{lawrence2004gaussian}
N.~Lawrence.
\newblock Gaussian process latent variable models for visualisation of high
  dimensional data.
\newblock In {\em NIPS}, 2004.

\bibitem{lawrence2005probabilistic}
N.~Lawrence.
\newblock Probabilistic non-linear principal component analysis with {G}aussian
  process latent variable models.
\newblock {\em JMLR}, 6(Nov):1783--1816, 2005.

\bibitem{liang2007infinite}
P.~Liang, S.~Petrov, M.~Jordan, and D.~Klein.
\newblock The infinite {PCFG} using hierarchical {D}irichlet processes.
\newblock In {\em EMNLP-CoNLL}, 2007.

\bibitem{liu2015real}
X.~Liu, A.~Nourbakhsh, Q.~Li, R.~Fang, and S.~Shah.
\newblock Real-time rumor debunking on twitter.
\newblock In {\em CIKM}, 2015.

\bibitem{ma2016detecting}
J.~Ma, W.~Gao, P.~Mitra, S.~Kwon, B.~J. Jansen, K.-F. Wong, and M.~Cha.
\newblock Detecting rumors from microblogs with recurrent neural networks.
\newblock In {\em IJCAI}, 2016.

\bibitem{ma2015detect}
J.~Ma, W.~Gao, Z.~Wei, Y.~Lu, and K.-F. Wong.
\newblock Detect rumors using time series of social context information on
  microblogging websites.
\newblock In {\em CIKM}, 2015.

\bibitem{ma2017detect}
J.~Ma, W.~Gao, and K.-F. Wong.
\newblock Detect rumors in microblog posts using propagation structure via
  kernel learning.
\newblock In {\em ACL}, 2017.

\bibitem{ma2018rumor}
J.~Ma, W.~Gao, and K.-F. Wong.
\newblock Rumor detection on twitter with tree-structured recursive neural
  networks.
\newblock In {\em ACL}, 2018.

\bibitem{mavroforakis2016modeling}
C.~Mavroforakis, I.~Valera, and M.~Gomez~Rodriguez.
\newblock Modeling the dynamics of online learning activity.
\newblock In {\em WWW}, 2017.

\bibitem{mcauliffe2008supervised}
J.~Mcauliffe and D.~Blei.
\newblock Supervised topic models.
\newblock In {\em NIPS}, 2008.

\bibitem{mccallum2005topic}
A.~McCallum, A.~Corrada-Emmanuel, and X.~Wang.
\newblock Topic and role discovery in social networks.
\newblock In {\em IJCAI}, 2005.

\bibitem{mimno2007expertise}
D.~Mimno and A.~McCallum.
\newblock Expertise modeling for matching papers with reviewers.
\newblock In {\em KDD}, 2007.

\bibitem{mimno2012topic}
D.~Mimno and A.~McCallum.
\newblock Topic models conditioned on arbitrary features with
  {D}irichlet-multinomial regression.
\newblock {\em UAI}, 2008.

\bibitem{mocanu2015collective}
D.~Mocanu, L.~Rossi, Q.~Zhang, M.~Karsai, and W.~Quattrociocchi.
\newblock Collective attention in the age of (mis) information.
\newblock {\em Computers in Human Behavior}, 51:1198--1204, 2015.

\bibitem{nguyen2018believe}
A.~Nguyen, A.~Kharosekar, S.~Krishnan, S.~Krishnan, E.~Tate, B.~Wallace, and
  M.~Lease.
\newblock Believe it or not: Designing a human-{A}{I} partnership for
  mixed-initiative fact-checking.
\newblock In {\em UIST}, 2018.

\bibitem{nguyen2012containment}
N.~P. Nguyen, G.~Yan, M.~T. Thai, and S.~Eidenbenz.
\newblock Containment of misinformation spread in online social networks.
\newblock In {\em WebSci}, 2012.

\bibitem{paisley2012discrete}
J.~Paisley, C.~Wang, and D.~Blei.
\newblock The discrete infinite logistic normal distribution.
\newblock {\em Bayesian Analysis}, 7(4):997--1034, 2012.

\bibitem{qazvinian2011rumor}
V.~Qazvinian, E.~Rosengren, D.~R. Radev, and Q.~Mei.
\newblock Rumor has it: Identifying misinformation in microblogs.
\newblock In {\em EMNLP}, 2011.

\bibitem{ramage2009labeled}
D.~Ramage, D.~Hall, R.~Nallapati, and C.~Manning.
\newblock Labeled {L}{D}{A}: A supervised topic model for credit attribution in
  multi-labeled corpora.
\newblock In {\em EMNLP}, 2009.

\bibitem{rapoza_2017}
K.~Rapoza.
\newblock Can ‘fake news’ impact the stock market, Feb 2017.

\bibitem{rashkin2017truth}
H.~Rashkin, E.~Choi, J.~Y. Jang, S.~Volkova, and Y.~Choi.
\newblock Truth of varying shades: Analyzing language in fake news and
  political fact-checking.
\newblock In {\em EMNLP}, 2017.

\bibitem{rosen2004author}
M.~Rosen-Zvi, T.~Griffiths, M.~Steyvers, and P.~Smyth.
\newblock The author-topic model for authors and documents.
\newblock In {\em UAI}, 2004.

\bibitem{roussas2014introduction}
G.~Roussas.
\newblock {\em An introduction to measure-theoretic probability}.
\newblock Academic Press, 2014.

\bibitem{seeger2003fast}
M.~Seeger, C.~Williams, and N.~Lawrence.
\newblock Fast forward selection to speed up sparse {G}aussian process
  regression.
\newblock In {\em AISTATS}, 2003.

\bibitem{sethuraman1994constructive}
J.~Sethuraman.
\newblock A constructive definition of {D}irichlet priors.
\newblock {\em Statistica Sinica}, 4:639--650, 1994.

\bibitem{snelson2006sparse}
E.~Snelson and Z.~Ghahramani.
\newblock Sparse {G}aussian processes using pseudo-inputs.
\newblock In {\em NIPS}, 2006.

\bibitem{tambuscio2015fact}
M.~Tambuscio, G.~Ruffo, A.~Flammini, and F.~Menczer.
\newblock Fact-checking effect on viral hoaxes: A model of misinformation
  spread in social networks.
\newblock In {\em WWW}, 2015.

\bibitem{teh2005sharing}
Y.~W. Teh, M.~Jordan, M.~Beal, and D.~Blei.
\newblock Sharing clusters among related groups: Hierarchical {D}irichlet
  processes.
\newblock In {\em NIPS}, 2005.

\bibitem{teh2008collapsed}
Y.~W. Teh, K.~Kurihara, and M.~Welling.
\newblock Collapsed variational inference for {HDP}.
\newblock In {\em NIPS}, 2008.

\bibitem{titsias2009variational}
M.~Titsias.
\newblock Variational learning of inducing variables in sparse {G}aussian
  processes.
\newblock In {\em AISTATS}, 2009.

\bibitem{titsias2010bayesian}
M.~Titsias and N.~Lawrence.
\newblock Bayesian {G}aussian process latent variable model.
\newblock In {\em AISTATS}, 2010.

\bibitem{tripathy2010study}
R.~M. Tripathy, A.~Bagchi, and S.~Mehta.
\newblock A study of rumor control strategies on social networks.
\newblock In {\em CIKM}, 2010.

\bibitem{tschiatschek2018fake}
S.~Tschiatschek, A.~Singla, M.~Gomez~Rodriguez, A.~Merchant, and A.~Krause.
\newblock Fake news detection in social networks via crowd signals.
\newblock In {\em WWW}, 2018.

\bibitem{vo2018rise}
N.~Vo and K.~Lee.
\newblock The rise of guardians: Fact-checking {U}{R}{L} recommendation to
  combat fake news.
\newblock In {\em SIGIR}, 2018.

\bibitem{vosoughi2018spread}
S.~Vosoughi, D.~Roy, and S.~Aral.
\newblock The spread of true and false news online.
\newblock {\em Science}, 359(6380):1146--1151, 2018.

\bibitem{wainwright2008graphical}
M.~Wainwright and M.~Jordan.
\newblock Graphical models, exponential families, and variational inference.
\newblock {\em Foundations and Trends{\textregistered} in Machine Learning},
  1(1--2):1--305, 2008.

\bibitem{wang2014mmrate}
S.~Wang, X.~Hu, P.~Yu, and Z.~Li.
\newblock {MMR}ate: inferring multi-aspect diffusion networks with
  multi-pattern cascades.
\newblock In {\em KDD}, 2014.

\bibitem{wu2015false}
K.~Wu, S.~Yang, and K.~Q. Zhu.
\newblock False rumors detection on sina weibo by propagation structures.
\newblock In {\em ICDE}, 2015.

\bibitem{wu2018tracing}
L.~Wu and H.~Liu.
\newblock Tracing fake-news footprints: {C}haracterizing social media messages
  by how they propagate.
\newblock In {\em WSDM}, 2018.

\bibitem{yang2012automatic}
F.~Yang, Y.~Liu, X.~Yu, and M.~Yang.
\newblock Automatic detection of rumor on sina weibo.
\newblock In {\em KDD}, 2012.

\bibitem{zhao2015enquiring}
Z.~Zhao, P.~Resnick, and Q.~Mei.
\newblock Enquiring minds: {E}arly detection of rumors in social media from
  enquiry posts.
\newblock In {\em WWW}, 2015.

\end{thebibliography}

\end{document}